\documentclass[10pt,prc,aps]{revtex4}
\draft
\usepackage{graphicx}
\usepackage{color}
\usepackage{mathtools}
\begin{document}

\title{The neutron skin of $^{48}$Ca and $^{208}$Pb: a critical analysis.
 }
\author{            
Francesca Sammarruca\footnote{ fsammarr@uidaho.edu }   }                                                        
\affiliation{ Physics Department, University of Idaho, Moscow, ID 83844-0903, U.S.A. 
}
\date{\today} 
\begin{abstract}
We discuss the neutron skins of $^{48}$Ca and $^{208}$Pb. We review and critically examine modern predictions and empirical constraints, with special attention to the different interpretations of the findings from the PREX-II experiment and the recently reported value of the neutron skin in $^{48}$Ca extracted from the CREX experiment. We argue that, in the spirit of the {\it ab initio} philosophy, the path to understanding the behavior of dense neutron-rich matter must not circumvent fundamental nuclear forces. Based only on that argument, a thick neutron skin in $^{208}$Pb is highly unlikely. 
\end{abstract}
\maketitle 

\section{Introduction} 
\label{Intro} 

The existence of the neutron skin is a remarkable feature of neutron-rich nuclei. Because neutrons don't bind, neutron excess, typically measured by the isospin asymmetry $ \alpha = (N-Z)/A$, is a destabilizing effect in a nucleus. As a consequence, some of the excess neutrons are ``pushed out" from the (neutron-enriched) core and form the skin. The same physics plays an important role in neutron stars, which are supported against gravitational collapse by the outward pressure existing in dense systems with high neutron concentration.  Studies of nuclear interactions in systems with high or extreme neutron to proton ratio are crucial for understanding the neutron driplines, the location of which is not well known. The new Facility for Rare Isotope Beams (FRIB), operational since May 2022, is expected to increase the number of known rare isotopes from 3000 to about 6000~\cite{frib}.

The role of microscopic nuclear physics is to predict observables based on fundamental nuclear forces derived from first principles. Comparison with measurements is then a true test of the predictive power of the theory. The neutron skin is not a ``conventional" observable, in that it's not measured directly. Instead, it is extracted from measurements of observables that are sensitive to the neutron density distribution in nuclei.  Naturally, the neutron density cannot be probed with electron scattering the way it's done for proton densities, and so different methods must be employed, such as experiments with hadronic probes or measurements which exploit the neutron {\it weak charge}. Those will be reviewed in Section~\ref{gen}.

 We recall that the neutron skin is defined as
\begin{equation}
S = <r^2>_n^{1/2} -  <r^2>_p^{1/2} \; ,
\label{skin}
\end{equation}
that is, the difference between the root mean square radii of the neutron and the proton distributions.

The common goal of experimentalists and theorists is to shed light on fundamental questions. In the case of neutron skins, the physics one wants to pin down concerns the aforementioned pressure that determines the spatial extension of the neutron skin, an information contained in the equation of state (EoS) of neutron-rich matter. The importance of this quantity cannot be overstated, given that its relevance extends from nuclei to compact astrophysical systems. These connections will be elucidated in Section~\ref{th}.

The purpose of this article is to review and critically examine modern predictions and empirical constraints as well as the different points of view currently being debated. A special focus is placed on the recently reported value of the neutron skin in $^{48}$Ca extracted from the CREX experiment~\cite{crex}, relative to the findings from the PREX-II~\cite{prexII} experiment.
 We end with thoughts and suggestions on the best way forward to strengthen the link between experiment and {\it ab initio} theory.

\section{Extracting neutron skins from experiments}
\label{gen}

\subsection{Brief review of useful facts}
\label{formu}

We will go through the main steps leading to the energy per nucleon in neutron-rich matter and related expressions, because they will be useful for the present discussion.
  We introduce the energy per nucleon, $e(\rho,\alpha)$, in an infinite system of nucleons at density $\rho$ and isospin asymmetry $\alpha = \frac{\rho_n - \rho_p}{\rho}$ -- namely, the EoS of neutron-rich matter -- and expand this quantity with respect to the isospin asymmetry parameter, $\alpha$:
\begin{equation}
e(\rho,\alpha) = e(\rho, 0) + \frac{1}{2} \Big ( \frac{\partial^2 e(\rho,\alpha)}{\partial \alpha^2} \Big )_{(\alpha = 0)}  \alpha^2 + \mathcal{O} (\alpha^{4})  \; .
\label{e_exp}
\end{equation} 
 Neglecting terms of order $\mathcal{O} (\alpha^{4})$, Eq.~(\ref{e_exp}) takes the well-known form:
\begin{equation}
e(\rho,\alpha) \approx e_0(\rho) + e_{sym}(\rho) \ \alpha^2  \; .
\label{asym_e}
\end{equation}
 Note that 
  $e_{sym}(\rho)$ = 
  $\frac{1}{2} \Big ( \frac{\partial^2 e(\rho,\alpha)}{\partial \alpha^2} \Big )_{\alpha = 0}$ and $e_0(\rho) =  e(\rho, 0)$, the EoS of isospin-symmetric nuclear matter.
Within the quadratic approximation  applied in Eq.~(\ref {asym_e}), the symmetry energy becomes the difference between the energy per neutron in neutron matter (NM) and the energy per nucleon in symmetric nuclear matter (SNM):
\begin{equation}
e_{sym} (\rho) = e_n (\rho) - e_0 (\rho) \; ,
\label{xxx}
\end{equation}
where $e_n(\rho) =  e(\rho, 1)$, the energy per neutron in pure NM.

We recall that $e_0(\rho)$ exhibits a minimum at a density approximately equal to the average central density of nuclei, $\rho_0$, a reflection of the saturating nature of the nuclear force. Next, we
expand the symmetry energy about the saturation point:
\begin{equation}
\label{yyy}
e_{sym} (\rho) \ \approx \ e_{sym} (\rho_{0}) + L \ \frac{\rho -\rho_{0}}{3 \rho_0} + \frac{K}{2} \frac{(\rho - \rho_{0})^2}{(3\rho_0)^2}  \; ,
\end{equation}
 where the expansion parameters are obviously related to the first and higher-order derivatives of $e_{sym}(\rho)$.
 $L$ is a measure of the slope of the symmetry energy at saturation:
\begin{equation}
\label{L}
L=3\rho_{0} \Big( \frac{\partial e_{sym}(\rho)}{\partial \rho} \Big)_{\rho_{0}}  \; .
\end{equation}
Furthermore, from Eqs.~(\ref{xxx}) and (\ref{L}), we see that $L$ is a measure of the slope of the NM EoS at saturation density, since the SNM EoS has
 a vanishing slope at that point.

Using the relation between pressure and energy density, we define the symmetry pressure:
\begin{equation}
\label{sym_pres}
P_{sym}(\rho) = \rho^2 \frac{\partial e_{sym}}{\partial \rho} = P_{NM}(\rho) - P_{SNM}(\rho) \; .
\end{equation}
If the derivative is evaluated at or very near $\rho_0$, the symmetry pressure is essentially the pressure in NM because the pressure in SNM vanishes at saturation. Then:
\begin{equation}
\label{ppp}
P_{NM}(\rho_0) = \Big (\rho^2 \frac{\partial e_n(\rho)}{\partial \rho} \Big )_{\rho_0} \; .
\end{equation}
From  Eq.~(\ref{L}) and Eq.~(\ref{ppp}), it is clear that the slope parameter $L$ is a measure of the pressure in NM around saturation density:
\begin{equation}
\label{L_P}
P_{NM}(\rho_0) = \rho_0 \frac{L}{3}  \; .
\end{equation}
It is then easy to understand how the symmetry energy slope, essentially a pressure gradient acting on excess neutrons, determines the formation and size of the neutron skin. Therefore, constraints on $L$ have the ability to provide constraints on the skin, and {\it vice versa}.     

\subsection{Experiments and phenomenological analyses} 
\label{exp}

Indirect measurements of the neutron skin in $^{208}$Pb and $^{48}$Ca have been performed using a variety of techniques, such as those listed below.  Parity-violating electron scattering will be addressed separately. Some representative 
measurements are:

\begin{itemize}

\item Proton-nucleus elastic scattering~\cite{Clark+2003, Zen+2010, Staro+1994, Shlo+1979};
\item Polarized proton-nucleus elastic scattering~\cite{Ray+1979, Zen+2018};
\item Pionic probes~\cite{Fried2012, Gib+1992};
\item Coherent pion photoproduction~\cite{Tarb+2014, Zana+2015};
\item Antiprotonic atom data~\cite{Bro+2007,Bro+2007_2, Trz+2001};
\item Electric dipole polarizability~\cite{Roca+2015,Roca+2013,Tam2011, Birk+2017};
\item Pygmy dipole resonsnces~\cite{Klim+2007};                             
\item Interaction cross sections with microscopic optical potentials~\cite{Mats+2022}; 
\item $(\alpha, \alpha')$ giant dipole resonance (GDR)~\cite{Kras+1994};
\item $\alpha$-particle scattering~\cite{Gils+1984};
\end{itemize}

Constraints directly on the symmetry energy and its density dependence have also been sought through a variety of techniques, such as:  data on nuclear masses across the periodic table~\cite{Mond+2015}, giant dipole resonance energies~\cite{Trip+2008}, electric dipole polarizability~\cite{Zhang+2015}, measurements of directed and elliptic flows in HI collisions~\cite{Russ+2016}, isobaric analog states~\cite{Dan+2014}, isospin diffusion in HI collisions~\cite{Tsang+2004}, neutron and proton transverse emission ratio measurements~\cite{Fam+2006},
HI collisions at intermediate energies~\cite{Yong+2020}.   

\begin{figure*}[!t] 
\centering
\hspace*{-1cm}
\includegraphics[width=8.7cm]{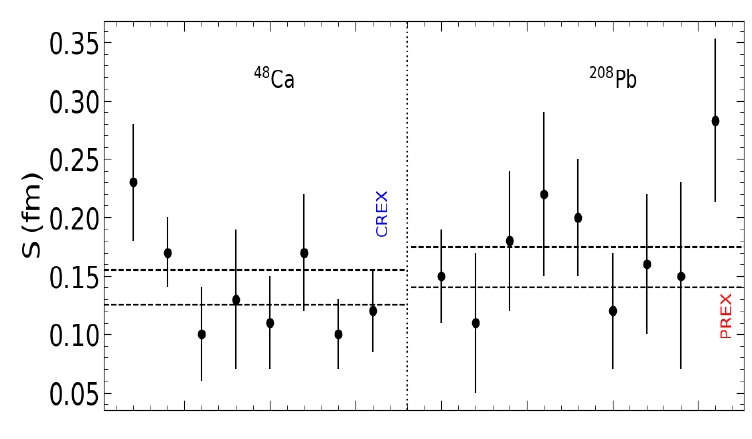}\hspace{0.01in} 
\vspace*{-0.5cm}
 \caption{(Color online) An overview of experimental constraints, taken from Ref.~\cite{Latt23} (see text for explanation). On the $^{48}$Ca ($^{208}$Pb) side, the last point on right is the result of CREX~\cite{crex} (PREX-II~\cite{prexII}). The horizontal lines mark the weighted means of the experiments $\pm$ one standard deviation, not including parity-violating electron scattering.
}
\label{Latt_fig}
\end{figure*}

In Table~\ref{tab_exp}, we 
summarize values for $^{208}$Pb and $^{48}$Ca neutron skins deduced from the indicated experiments. There is a considerable spread, as the result of a multitude of methods and theoretical input over decades. Analyses of hadronic scattering experiments, in particular, require modeling of the nuclear potential. Based on previous measurements of the skin in $^{48}$Ca, we see no strong reasons to deem the CREX result surprising or unexpected, whereas the opposite is true for $^{208}$Pb. These observations are well captured in Fig.~\ref{Latt_fig}, the content of which we have extracted from Fig.~8 of  Ref.~\cite{Latt23}.  On the Calcium side, from left to right, the data points are from Refs.~\cite{Ray+1979},~\cite{Zen+2018},~\cite{Clark+2003},~\cite{Fried2012},~\cite{Gib+1992},~\cite{Gils+1984},~\cite{Shlo+1979}. On the Lead side, in the same order, the first two points correspond to Refs.~\cite{Tarb+2014} and~\cite{Fried2012}. The third point is at 0.18 $\pm$ 0.06 fm, as in Ref.~\cite{Latt23}, whereas, from the same references,~\cite{Bro+2007, Bro+2007_2}, we read the values reported in Table~\ref{tab_exp}. The remaining points are from Refs.~\cite{Zen+2010},~\cite{Staro+1994},~\cite{Clark+2003},~\cite{Ray+1979},~\cite{Fried2012}.
A comment is in place with regard to the values from Ref.~\cite{Clark+2003}, 3$^{rd}$ and 6$^{th}$ data points, at 0.098 $\pm$ 0.043 fm and 0.119 $\pm$ 0.045 fm, for $^{48}$Ca and $^{208}$Pb, respectively. From Ref.~\cite{Clark+2003}, we read the smaller values shown in Table~\ref{tab_exp}.

For $^{48}$Ca ($^{208}$Pb),  the last point on the right is the result of CREX~\cite{crex} (PREX-II~\cite{prexII}).

\begin{table*}[t]
\caption{Values of the neutron skins in $^{48}$Ca and in $^{208}$Pb from a variety of experimental methods.  }
\label{tab_exp}
\begin{tabular*}{\textwidth}{@{\extracolsep{\fill}}ccccc}
\hline
\hline
 Type of measurement & Extracted neutron skin in $^{48}$Ca & Extracted neutron skin in $^{208}$Pb  \\ 
\hline 
\hline
Proton-nucleus scattering~\cite{Clark+2003}  & 0.056 - 0.102 & 0.083 - 0.111 \\ 
 Proton-nucleus scattering~\cite{Zen+2010} &       &  0.211$^{+0.054}_{-0.063}$   \\
Proton-nucleus scattering~\cite{Staro+1994} &    &  0.20 $\pm$ 0.04  \\
Proton-nucleus scattering~\cite{Shlo+1979} &   0.10 $\pm$ 0.03 &      \\
Polarized proton-nucleus scattering~\cite{Ray+1979}  & 0.23 $\pm$ 0.05   &  0.16 $\pm$ 0.05  \\
Polarized proton-nucleus scattering~\cite{Zen+2018}  & 0.168$^{+0.025}_{-0.028}$   &    \\   
Polarized proton-nucleus scattering~\cite{Zen+2010}  &    &   0.211 $^{+0.054}_{-0.063}$  \\              
Pionic probes~\cite{Fried2012}    &0.13 $\pm$ 0.06   &  0.11 $\pm$ 0.06 \\
Pionic probes~\cite{Gib+1992}    &0.11 $\pm$ 0.04   &       \\
Coherent $\pi$ photoproduction~\cite{Tarb+2014} &     &  0.15 $\pm$ 0.03$^{+0.01}_{-0.03}$   \\
Coherent $\pi$ photoproduction~\cite{Zana+2015} &     &  0.20 $^{+0.01}_{-0.03}$   \\ 
Antiprotonic atoms~\cite{Bro+2007} &       &   0.20 ($\pm$ 0.04) ($\pm$ 0.05)    \\
Antiprotonic atoms~\cite{Bro+2007_2} &       &   0.16 ($\pm$ 0.02) ($\pm$ 0.04)    \\
Antiprotonic atoms~\cite{Trz+2001} &       &   0.15 $\pm$ 0.02    \\                         
Electric dipole polarizability~\cite{Roca+2015} &    &  0.13 - 0.19 \\
Electric dipole polarizability~\cite{Roca+2013} &    &  0.165 ($\pm$0.09)($\pm$ 0.013) ($\pm$0.021) \\
Electric dipole polarizability &      &             \\
{\it via} polarized scattering at forward angle~\cite{Tam2011} &    &  0.156 $^{+0.025}_{-0.021}$ \\     
Electric dipole polarizability~\cite{Birk+2017} & 0.14 - 0.20   &   \\
Pygmy dipole resonances~\cite{Klim+2007} &      & 0.18 $\pm$ 0.035   \\
Interaction cross sections~\cite{Mats+2022} &   0.105 $\pm$ 0.06 &       \\
$(\alpha, \alpha')$ GDR 120 MeV~\cite{Kras+1994} &       & 0.19 $\pm$0.09  \\
$\alpha$-particle scattering~\cite{Gils+1984}  & 0.171 $\pm$ 0.05  &        \\
\hline
\hline
\end{tabular*}
\end{table*}

\begin{figure*}[!t] 
\centering
\hspace*{-1cm}
\includegraphics[width=7.7cm]{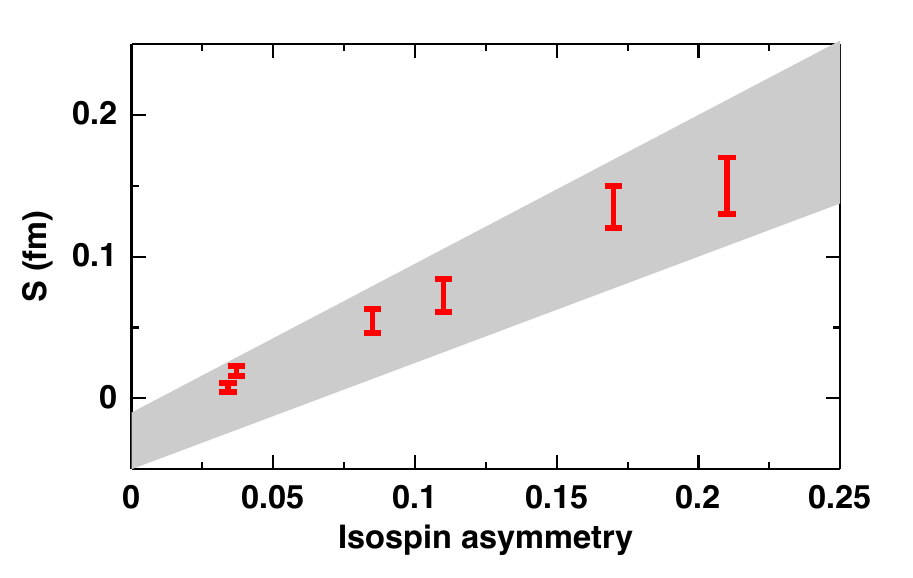}\hspace{0.01in} 
\vspace*{-0.1cm}
 \caption{(Color online) 
Red bars: Neutron skin of $^{58}$Ni, $^{27}$Al, $^{59}$Co, $^{90}$Zr, $^{48}$Ca,
and $^{208}$Pb, in order of increasing isospin asymmetry, from Ref.~\cite{FS_2022}. The shaded area is bounded by linear fits to the data~\cite{Swia+2005}.
}
\label{fs_fig}
\end{figure*}

We end this section with Fig.~\ref{fs_fig}, showing our predictions of the neutron skin for selected nuclei as a function of the isospin asymmetry~\cite{FS_2022}, confirming a nearly linear relation between these two quantities and, thus, motivating the expectation that the skins of $^{48}$Ca and $^{208}$Pb should be close. The boundaries of the grey area are linear fits to the data from Ref.~\cite{Swia+2005}.

\begin{figure*}[!t] 
\centering
\hspace*{-1cm}
\includegraphics[width=7.7cm]{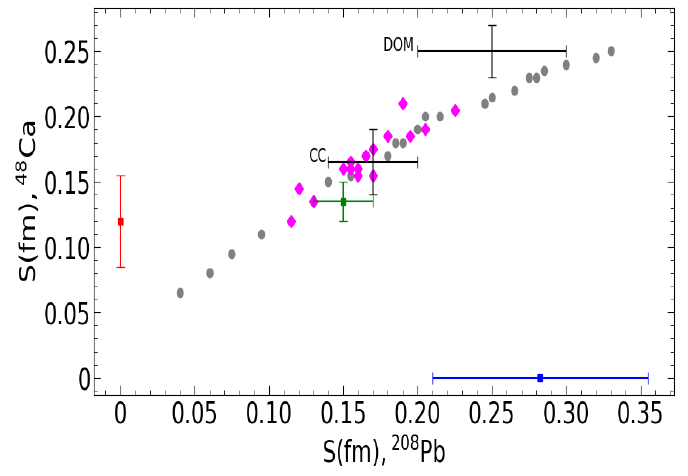}\hspace{0.01in} 
\vspace*{-0.05cm}
 \caption{(Color online) The neutron skin of $^{48}$Ca {\it vs.} the one of $^{208}$Pb. The red and blue bars are the results from CREX and combined PREX-I and PREX-II, respectively. The gray circles and pink diamonds are the result of relativistic and non-relativistic mean-field models, respectively.
Coupled cluster
and dispersive optical model predictions are indicated as CC and DOM, respectively.
}
\label{rrr}
\end{figure*}

\subsection{Parity-violating electron scattering}

The parity-violating electron scattering asymmetry, $A_{PV}$, is defined for a spin-zero nucleus as
\begin{equation}
\label{apv}
A_{PV} = \frac{\sigma_R - \sigma_L}{\sigma_R + \sigma_L} \; ,
\end{equation}
where $\sigma_{R(L)}$ is the elastic cross section for right (left) handed electrons~\cite{Donn+1989}.
 $A_{PV}$ is proportional to the ratio of weak ($F_W(q)$) to charge ($F_{ch}(q)$) form factors, whith 
$q$ the four-momentum transfer. $F_{ch}(q)$ is taken from 
existing measurements and $F_W(q)$ is  extracted from the measured $A_{PV}$. We recall that the weak and charge form factors are the Fourier transforms of the weak charge density and the charge density, respectively:
\begin{equation}
\label{fw}
F_W(q) = \frac{1}{Q_W} \int d^3r j_0(qr) \rho_W(r) \; ,
\end{equation}
where $Q_W$ is the weak charge of the nucleus, and 
\begin{equation}
\label{fch}
F_{ch}(q) = \frac{1}{Z} \int d^3r j_0(qr) \rho_{ch}(r) \; .
\end{equation}
In Eq.~(\ref{fw}), a form is assumed for $\rho_W(r)$ and the radius parameter of the density function is adjusted to reproduce the experimental $A_{PV}$. The CREX result is found insensitive to the assumed form for the weak charge density~\cite{crex}.

\section{The status of {\it ab initio} theory}
\label{th}

\subsection{Development of microscopic nuclear forces}

Our still incomplete knowledge of nuclear forces is the result of decades of struggle. Currently, the optimal approach to the construction of nuclear forces is based on the understanding that 
the energy scale determines the appropriate degrees of freedom of the theory -- the central
concept to the development of chiral effective field theory (EFT)~\cite{Wei90,Wei92}. Here, we provide only a brief summary of the strongest features of chiral EFT.

Chiral EFT allows the development of nuclear interactions as an expansion where theoretical uncertainties can be assessed at each order.  The organizational scheme that controls the expansion is known as
``power counting."  The crucial point is that
chiral EFT maintains consistency with the underlying fundamental theory of strong interactions, quantum chromodynamics (QCD), through the symmetries and
symmetry breaking mechanisms of the low-energy QCD Lagrangian. 
The first step towards the development of an EFT is the identification of a ``soft scale'' and a ``hard scale," which is suggested by the hadron spectrum, observing the large separation between the mass of the pion and the mass of the vector meson $\rho$. It is therefore natural to identify the pion mass and the $\rho$ mass (approximately 1 GeV) with the soft and the hard scale, respectively. Moreover, since quarks and gluons are ineffective degrees of freedom in the low-energy regime, pions and nucleons can be taken as the appropriate degrees of freedom of the EFT.
Having identified pions and nucleons as the appropriate degrees of freedom of the EFT, one can proceed to construct the Lagrangian of the effective theory:
\begin{equation}
\label{lagr}
\mathcal{L}_{eff} = \mathcal{L}_{\pi\pi} + \mathcal{L}_{\pi N} + \mathcal{L}_{NN} + ...    \; ,
\end{equation}
which is then expanded in terms of a natural parameter, identified with the ratio of the ``soft scale'' over the ``hard scale,''  $\frac{Q}{\Lambda_{\chi}}$. $Q$ is of the order of the pion mass, whereas $\Lambda_{\chi}$ is the energy scale of chiral symmetry breaking, approximately 1 GeV.
 The contributions to the effective Lagrangian are arranged according to the power counting scheme, with increasing order resulting in smaller terms. While the expansion itself is, of course, infinite, at each order we are assured that the number of terms is finite and the contributions well defined.
The combination of meson-theoretic NN potentials augmented with selected 3NF is an outdated paradigm, from which it's essentially impossible to estimate the theoretical uncertainty of a prediction. 

 With chiral EFT, there has been enormous progress in the understanding and development of few-nucleon forces. We are on the right path, but there is still much to accomplish.
Convergence at N$^3$LO needs to be on more robust grounds. Furthermore, there are indications that  specific components of the 3NF at N$^4$LO have the potential to solve some outstanding problems in microscopic nuclear structure~\cite{Mac2023}.

\subsection{Neutron skin predictions}

\begin{table*}[t]
\caption{Status of {\it ab initio} predictions for the skins (in fm) in $^{48}$Ca and in $^{208}$Pb. The two results from Ref.~\cite{Hu+2022} for both $^{48}$Ca and in $^{208}$Pb show acceptable ranges within 68\% and 90\% of the credibility region (CR).}
\label{tab_ab}
\begin{tabular*}{\textwidth}{@{\extracolsep{\fill}}cccc}
\hline
\hline
 Nucleus & Predicted skin &  Source   \\ 
\hline 
\hline
 $^{48}$Ca &  0.120 - 0.150 & Ref.~\cite{Hag+2016}       \\
$^{48}$Ca &  0.141 - 0.187 & Ref.~\cite{Hu+2022}   68\% CR   \\
$^{48}$Ca &  0.123 - 0.199 & Ref.~\cite{Hu+2022}    90\% CR  \\
$^{208}$Pb &  0.139 - 0.200 & Ref.~\cite{Hu+2022}  68\% CR     \\
$^{208}$Pb &  0.120 - 0.221 & Ref.~\cite{Hu+2022}   90\% CR   \\
$^{48}$Ca &  0.114 - 0.186 & Ref.~\cite{Nov+2023}       \\
$^{208}$Pb &  0.184 - 0.236 & Ref.~\cite{Nov+2023}      \\
$^{48}$Ca &  0.12 - 0.15 & Ref.~\cite{Sam2022}       \\
$^{208}$Pb &  0.13 - 0.17 & Ref.~\cite{Sam2022}   \\
\hline
\hline
\end{tabular*}
\end{table*}

We show in Table~\ref{tab_ab} recent predictions for the skins in $^{48}$Ca and in $^{208}$Pb. Some comments are in place. In both Ref.~\cite{Hag+2016} and Ref.~\cite{Nov+2023}, the nature of the NN chiral potentials, N$^2$LO$_{sat}$ and $\Delta$N$^2$LO$_{GO}$, is such that the results are not truly {\it ab initio}.
Note, also, that the value for $^{208}$Pb from Ref.~\cite{Nov+2023} is not a prediction, but was obtained using the linear regression the authors constructed from the skins of lighter nuclei.

Figure~\ref{rrr} shows the neutron skin in $^{48}$Ca {\it vs.} the one in $^{208}$Pb. All values, except for the green bars, have been extracted from Fig.~5 of Ref.~\cite{crex}. The PREX-II and PREX-I combined experimental result is shown by the blue bar, while the red vertical bar is the CREX result. The
gray circles (pink diamonds) show results from a variety of relativistic
(non-relativistic) density functionals, which give values of the $L$ parameter ranging from small and negative to large and positive. 
Coupled cluster (CC)~\cite{Hag+2016, Nov+2023}
and dispersive optical model (DOM) predictions~\cite{Atk+2020}  are also displayed. Our predictions are shown by the green bars~\cite{Sam2022}. There exist, of course, relativistic mean-field (RMF) models that agree with the PREX result, and, correspondingly, generate values for $^{48}$Ca that are also on the larger side.

We conclude that a value between 0.212 fm and 0.354 fm (0.283 $\pm$ 0.071) for the skin of $^{208}$Pb is outside the boundaries set by microscopic theory. Simultaneous consistency with both CREX and PREX seems to be a challenge even for phenomenology.
On the other hand, PREX aside, a small skin for $^{48}$Ca does not appear to be peculiar based on the facts we reviewed above, see also Fig.~\ref{Latt_fig}.

\section{Further Discussion}

Irrespective of the inconsistency between PREX and CREX, there are far-reaching questions one must consider with regard to a thick skin in $^{208}$Pb and its ramifications.

\subsection{Direct Urca processes}

We define the total energy per baryon in $\beta$-equilibrated matter (in absence of muons) as 
\begin{equation}
\label{total_eng_per_prt}
\begin{aligned}
e_T (\rho, Y_p) = e_0(\rho) + e_{sym} \ (1 -2 Y_{p})^2 + 
 e_e + \sum_{i = n,p} Y_i\cdot m_i  \; ,
\end{aligned}
\end{equation}
where $Y_{p}$ is the proton fractions. The last term accounts for the baryon rest masses  (in units of energy), while $e_{e}$ is the electron energy. The particle fractions are 
\begin{equation}
\label{frac_expr}
Y_i=\frac{\rho_i}{\rho}  \; ,
\end{equation}
and the chemical potentials are given by:
\begin{equation}
\label{chempot_frac}
\mu_{i} = \frac{\partial \epsilon_{i}}{\partial \rho_{i}} = \frac{\partial e_{i}}{\partial Y_{i}}  \; .
\end{equation}
Naturally, we impose the constraints of fixed baryon density, Eq.~(\ref{frac_bary}), and global charge neutrality, Eq.~(\ref{frac_chrg}): 
\begin{equation}
\label{frac_bary}
\rho_p + \rho_n = \rho \quad \Rightarrow \quad Y_p + Y_n = 1  \; ,
\end{equation}
\begin{equation}
\label{frac_chrg}
\rho_p = \rho_e  \quad \Rightarrow \quad Y_p = Y_e  \; .
\end{equation}
The electron energy is easily written as
\begin{equation}
\label{epp_el}
e_{e} = \frac{\hbar c}{4\pi^2} \ \rho^{\frac{1}{3}} (3\pi^2 Y_{p})^{\frac{4}{3}} \; ,
\end{equation}
and thus 
\begin{equation}
\label{el_mu_chempot}
\begin{aligned}
\frac{\partial e_e}{\partial Y_{p}} = \frac{\hbar c}{3\pi^2} \ (3\pi^2)^{\frac{4}{3}} (\rho \ Y_{p})^{\frac{1}{3}} \; .
\end{aligned}
\end{equation}
Noting that 
\begin{equation}
\label{yyyy}
\frac{\partial e_p}{\partial Y_{p}} = \frac{\partial e_n}{\partial Y_{n}} - \frac{\partial e_e}{\partial Y_{e}} \; ,
\end{equation}
and using Eqs.~(\ref{total_eng_per_prt}) and~(\ref{yyyy}), we obtain:
\begin{equation}
\label{esym_x}
\begin{aligned}
4 \ e_{sym} (\rho) \ (1 - 2\cdot Y_p) = \hbar c (3\pi^2)^{\frac{1}{3}} (\rho \ Y_{p})^{\frac{1}{3}} \; ,
\end{aligned}
\end{equation}
where we have neglected the mass difference between the proton and the neutron.

If $\rho_{DU}$ is the density at which $Y_p$ is equal to the value needed for direct Urca processes (DU), about $1/9$, the following relation holds:
\begin{equation}
\label{urca2}
\begin{aligned}
e_{sym} (\rho_{DU}) = \hbar c \frac{9}{28} (\pi^2/3)^{\frac{1}{3}}  (\rho_{DU})^{1/3} \; .
\end{aligned}
\end{equation}
This simple relation can be quite insightful.
If, for example, $\rho_{DU}$ is close to saturation density, the symmetry energy would be over 50 MeV at that point. If $\rho_{DU}  = \frac{2}{3}\rho_0$, the symmetry energy at that density would be over 40 MeV. In microscopic predictions, DU processes are more likely to open at a few to several times normal density (based on projections, since chiral EFT predictions cannot be extended to such high densities). Using PREX II constraints, the Urca threshold is found to be approximately 1.5$\rho_0$ or just above 0.2 $fm^{-3}$~\cite{Kumar+}. From Eq.~(\ref{urca2}), the value of the symmetry energy at 1.5$\rho_0$ is then about 58 MeV, clearly indicationg an unusually steep density dependence, taking the PREX II value of 38.1 $\pm$ 4.7 MeV at saturation. 

In Ref.~\cite{Kumar+}, the authors develop different parametrizations of RMF models constrained by CREX results, PREX II results, or a combination of both, and report that the direct Urca threshold density decreases from 0.71 fm$^{-3}$ to 0.21 fm$^{-3}$ as the skin of $^{208}$Pb increases from 0.13 fm to 0.28 fm, with expected implication for cooling processes.
To the best of our knowledge, there is no evidence for rapid neutrino cooling {\it via} direct Urca for low-mass (low central density) neutron stars. In fact, from analyses including luminosities and ages determined from observations of isolated neutron stars~\cite{Belo+2019}, one may conclude that the direct Urca process, although possible at the central densities of neutron stars with masses between 1.7 and 2.0 solar masses, is unlikely around $M = 1.7 M_{\odot}$, and is likely, but not the principal cooling mechanism, for the $M = 2.0 M_{\odot}$ stars.

\subsection{Impact of the isovector component of the free-space nuclear force}

The importance of a realistic isovector component of the nuclear force (that is, carefully calibrated through free-space NN data), on the density dependence of the symmetry energy has been demonstrated~\cite{Hu+2022, FS_2023}. Relaxing the constraint of accurate phase shifts for the isospin-1 $S$ and $P$ waves leads to drastic variations of the pressure in NM~\cite{FS_2023} and, consequently, the neutron skin.

For the purpose of the present discussion, it's useful to recall that, typically, the isovector sector of RMF functionals includes coupling of the nucleon to the isovector $\rho$-meson and a nonlinear
$\omega - \rho$ isoscalar-isovector cross-coupling term. In the models of Ref.~\cite{Hor+2023}, designed to reconcile the PREX and CREX results, the novel feature is the inclusion of the scalar isovector $\delta$-meson (better known as the $a_0$-meson).  The isospin splitting resulting from the isovector nature of the new contribution opens an additional fitting degree of freedom~\cite{Hor+2023}. 

 As mesons heavier than the pion do not enter the development of the nucleon-nucleon (NN) force in chiral EFT, we will discuss this important point from the perspective of quantitative one-boson-exchange potentials (OBEP), as we already did over 10 years ago~\cite{FS_2011}. 
Studies of the $\rho$ and $\delta$ contributions to the potential part of the symmetry energy in quantum hydrodynamics (QHD) models can be found in Refs.~[30-35] of Ref.~\cite{Hor+2023}, where those contributions are shown to be opposite in sign and very large in magnitude around normal density. 
Thus, in QHD-inspired models, the interplay between $\rho$ and $\delta$ is treated as the equivalent, in the isovector channel, of the interplay between $\sigma$ and $\omega$ in the isoscalar channel. In contrast, the role of the $\delta$-meson in meson theory is subtle (although important) and it's seen in the difference of its contributions to isospin-1 or isospin-0 partial waves, particularly $^1S_0$ and $^3S_1$.
The dramatic difference between the description of the isovector channel in QHD-based models or realistic meson models on originate from several sources, including the absence of the pion in QHD, and the fact that meson contributions in a microscopic approach are iterated, and thus reduced by Pauli blocking.

In Ref.~\cite{FS_2011}, we have shown the difference  between the potential energy contributions to NM and SNM from the isovector mesons, to estimate the effect of each meson on the potential energy part of the symmetry energy. The impact of the pion
on the symmetry energy is by far the largest. On the other hand, mean-field theories are generally pionless,
because the bulk of the attraction-repulsion balance needed for a realistic description
of nuclear matter can be technically obtained from $\sigma$ and $\omega$ only, an observation that is at the very foundation of the
Walecka model~\cite{Serot+1997}. On the other hand, in any theory of nuclear forces, the pion is the most
important ingredient. Chiral symmetry is spontaneously broken in low-energy QCD and the pion emerges as the
Goldstone boson of this symmetry breaking.  Moreover, NN scattering data, and most definitely the deuteron, cannot be described without the pion. 

In conclusion, developing additional density functionals with unrealistic isovector contributions for the purpose of reconciling the parity-violating experiments does not provide more clarity. The road to understanding the behavior of dense neutron-rich matter must go through {\it ab initio} nuclear forces.

\section{Conclusions}

The above has been a critical analysis of the available facts on the neutron skins in $^{208}$Pb and $^{48}$Ca , from both the experimental and the theoretical side. 

We have pointed out and discussed problematic aspects that arise in conjunction with a stiff EoS and the corresponding
  thick skin in $^{208}$Pb.  We emphasized that such thick skin is outside the boundaries of {\it ab initio} nuclear theory.

On the other hand, in the context of existing measurements and microscopic predictions, the CREX result for the skin of $^{48}$Ca does not appear to be an outlier.

We wait with excitement for the  MREX experiment at the MESA accelerator, which promises to measure the neutron skin of $^{208}$Pb with {\it ultimate precision}~\cite{mrex}.

 Laboratory neutron skin measurements have important implications for neutron star
properties, such as radius, tidal deformability, and cooling processes, and are 
complementary to astronomical observations. In the ``multi-messanger" era, it is especially important to keep a broad view and, at the same time, stay in close touch with first principles, as we study the remarkable connection between microscopic physics and ``telescope physics" enabled by the equation of state.

\section*{Acknowledgments}
This work was supported by 
the U.S. Department of Energy, Office of Science, Office of Basic Energy Sciences, under Award Number DE-FG02-03ER41270.


\end{document}